\documentclass[twocolumn,showpacs,preprintnumbers,amsmath,amssymb]{revtex4}

\usepackage{graphicx}
\usepackage{dcolumn}
\usepackage{bm}

\begin{document}

\title{Insecurity of position-based quantum cryptography protocols against entanglement attacks}

\author{Hoi-Kwan Lau\footnote{kero.lau@utoronto.ca} and Hoi-Kwong Lo\footnote{hklo@comm.utoronto.ca}}
 \affiliation{Center for Quantum Information and Quantum Control (CQIQC),
 Department of Physics and Department of Electrical and Computing Engineering,
 University of Toronto, 60 Saint George Street, Toronto, M5S 1A7, Ontario, Canada}

\date{\today}\begin{abstract}
Recently, position-based quantum cryptography has been claimed to be unconditionally secure. On the contrary, here we show that the existing proposals for position-based quantum cryptography are, in fact, insecure if entanglement is shared among two adversaries.  Specifically, we demonstrate how the adversaries can incorporate ideas of quantum teleportation and quantum secret sharing to compromise the security with certainty.  The common flaw to all current protocols is that the Pauli operators always map a codeword to a codeword (up to an irrelevant overall phase).  We propose a modified scheme lacking this property in which the same cheating strategy used to undermine the previous protocols can succeed with a rate at most $85\%$.  We conjecture that the modified protocol is unconditionally secure and prove this to be true when the shared quantum resource between the adversaries is a two- or three- level system.
\end{abstract}



\maketitle

\section{\label{intro}Introduction}

Quantum cryptography has both power and limitations. Whereas quantum cryptography can offer unconditional security \cite{Mayers:2001p6348, Lo:1999p6069, Shor:2000p6087} in communications through quantum key distribution (QKD) \cite{BENNETT:1984p6499, Ekert:1991p6454} and secure multi-party computations through quantum secret sharing \cite{Hillery:1999p5705, Cleve:1999p5862}, it cannot protect private information in secure two-party computations due to standard no-go theorems in quantum bit commitment \cite{Lo:1997p5814, Mayers:1997p5812} and quantum oblivious transfer \cite{Lo:1997p6194}.
So, what is the boundary to the power of quantum cryptography?

To answer the above question, there has been much research interest in the subject of quantum coin flipping.  See, for example, \cite{Chailloux:2009p6335} and references cited therein.  In this paper, we focus on another proposed application of quantum cryptography. It is called position-based cryptography (PBC) \cite{Chandran:2009p5809}. The goal of position-based cryptography is for a prover to prove to a set of cooperating
spatially separated verifiers that he is at (or in the small neighborhood of) a particular spatial location.

Why is position-based cryptography interesting?
In everyday life, we tend place trust on spatial locations.
For instance, when we go to a branch of a bank to deposit some money,
we seldom ask a teller to show his/her identity to us to prove
that he/she is indeed a bank employee, rather than someone faking
as one. Why? This is because, for instance
we went to the branch yesterday and knew where it was and so, the fact
that a person is today standing
in a particular location in a branch of a bank convinces us that
he/she should be trusted as a bank employee.

Position-based cryptography might also be of practical interest
in, for example, automatic road tolling in vehicular communication systems \cite{website:SVC}.
Instead of collecting road tolls manually or installing many automatic
collection stations in each highway entrance and exit, it would be nice to
use, for example, satellites to track all vehicles in a highway and
charge them road tolls automatically, according to the paths taken.
For such a road toll system to be fool-proof, it is important to
ensure that a cheater cannot fool the verifiers of his whereabout.

Unfortunately, in the classical world, unconditionally secure
position-based cryptography has been proven to be impossible
\cite{Chandran:2009p5809, Chandran:2010p5507}.  The reason is that classical message
can be cloned by cheaters perfectly who can resend a copy to the
authorized receiver. So, both senders and receivers cannot detect
an intercept-and-broadcast attack.

Quantum cryptography has a fundamental advantage over classical cryptography due to the quantum no-cloning theorem \cite{WOOTTERS:1982p6285, Dieks:1982p6286}.
In view of the success in quantum key distribution, it is an interesting question
to ask whether PBC can be implemented with unconditionally security under quantum setting.  
As far as we know, the possibility of position-based quantum cryptography (PBQC) was first
studied by Kent under the name of `quantum tagging' as early as 2002.  Based on the idea, a patent of quantum tagging system
introduced by Kent \textit{et al.} was granted in 2006 \cite{patent:QT}. However,
their results have not appeared in the academic literature until 2010 \cite{Kent:2010p5759}. 
Recently, before the appearance of \cite{Kent:2010p5759}, two PBQC protocols have been independently proposed by Chandran \textit{et al.} \cite{Chandran:2010p5507} (hereafter denoted as \textit{Protocol A}), and
Malaney \cite{Malaney:2010p5361, Malaney:2010p5555} (hereafter denoted as \textit{Protocol B}).
Protocol A is claimed to be unconditionally secure with a detailed proof of security based on complementary information tradeoff argument. Protocol B is also claimed to be unconditionally secure due to the quantum no-cloning theorem,
but no detailed security proof has been given.

Contrary to the claims of unconditional security, both protocols are, in fact, insecure.  Cheaters can make use of entanglement to conduct non-local operations to produce the same response as the honest case.  Independent of our present work, Kent, Munro, and Spiller \cite{Kent:2010p5759} have discussed some conditions required for a secure PBQC protocol.  They address that several types of PBQC scheme are insecure against teleportation-based attacks. They outline attacks if the locations of reference stations and authorized receiver are collinear. Their attack applies to Protocols A and B
for the case of one spatial dimension.

There are two objectives in this paper. First, we show how existing PBQC protocols (Protocols A and B)
can be cheated by using entangled resources, and discuss why the protocols are insecure. We discuss
not only the case of one spatial dimension, but also higher dimensions.
Second, knowing the reasons, we propose a modified protocol and discuss its security.
Our paper is organized as follows.  In Section \ref{protocol}, we outline the procedure of both Protocol A and Protocol B.  In the following Section \ref{cheat 1D}, we consider the case where the number of reference stations $N=2$.  Similar to \cite{Kent:2010p5759} but, more explicitly and in a step-by-step manner, we show how the protocols
can be cheated with certainty.  In Section \ref{cheat 3D}, we consider the cases where $N>2$, which can be cheated by techniques of quantum secret sharing and cluster state quantum computation.   
%Contrary to the apparent belief of various authors that $N>2$ cases can be cheated by simply extending the teleportation-based attacks in $N=2$ case, techniques beyond quantum teleportation, such as quantum secret sharing and cluster state quantum computation, are required.  
The reason of the insecurity of both protocols and the loophole of the claimed security proof are discussed in
Section \ref{cheating principle}.  In Section \ref{proposal}, we give our modified protocol and examine its security under our cheating scheme.  Our protocol is proved to be secure in Section \ref{security} if cheaters share entangled qubits or qutrits only.  Finally we summarize our paper in Section \ref{conclusion} with brief discussions.

\section{PBQC protocols\label{protocol}}

%An effective PBQC protocol ought to divide the quantum information `disunite' enough, such that cheaters cannot gain adequate description about the quantum system for a perfect measurement.  To the best of our knowledge, there are two PBQC protocol proposed so far, Protocol A and Protocol B.
%However, we find that both protocol are not robust against teleportation-assisted attack.  In this section, we first outline the two protocols, then demonstrate our attack in 1-dimensional, and finally generalize it to higher dimension (more stations).

Here, for simplicity,
we assume that all honest parties have synchronized clocks and work with flat Minkowski space-time
in special relativity.
The idea of position-based quantum cryptography is to divide encoded quantum information into several parts (but possibly entangled) and distribute to $N$ reference stations $V_1,\ldots,V_N$ at various separated locations.  The divided pieces are then sent from different directions to an authorized receiver $P$, who is located at a pre-assumed position $\vec{x}_P$ surrounding by a finite secure region that no cheaters can get into.  The interval of sending is well chosen in such a way that $\vec{x}_P$ is the unique position where the shortest traveling time is needed for all pieces of information to come together.  For simplicity, we hereafter assume $\vec{x}_P$ locates at midway between reference stations (or equidistance position in higher dimension scenario) and the divided information are sent simultaneously.  Measurement is immediately conducted by $P$ on the quantum system, and the result is broadcasted for verification.  The intuition is as follows. Since perfect quantum measurement (acquire complete status about quantum particle) can only be achieved with adequate knowledge about the system, such as the correct basis of quantum state, cheaters outside $\vec{x}_P$ ought to wait for a longer time for enough information to conduct perfect measurement. Otherwise, they are only able to conduct imperfect measurements.  Therefore, references can authenticate receiver's position by checking the response time and error rate of broadcasted measurement results.  Procedures of two existing proposals are outlined as follows.

\subsection{Protocol A}

%\subsubsection{Protocol}

The idea of Protocol A is to send the basis of measurement and the encoded qubit separately from different reference stations \cite{Chandran:2010p5507}.  Security of this protocol was supposed to rely on the idea that quantum system can be measured perfectly only if the correct measurement basis is obtained.  Explicit procedures of Protocol A follows.

\textit{Step 1.} Station $V_1$ encodes a message $u \in \{0,1\}$  as a qubit $|u \rangle$, where $|0\rangle$ and $|1\rangle$ are $+1$ and $-1$ eigenstates of Pauli $Z$ operator respectively.  Inspired by the well-known BB84 QKD protocol, $V_1$ encrypts the message by applying the transformation $H^q$ on the qubit, where $H$ is the Hadamard gate, $q$ is a random bit valued $0$ or $1$.

\textit{Step 2.} $V_1$ generates $N-2$ random bits $q_2, q_3, \ldots, q_{N-1}$, and decide a bit $q_N$ by the relation
\begin{equation}
q=q_2+q_3+\dots+q_N \mod 2~.
\end{equation}
The bits $q_2, q_3, \ldots, q_N$ are distributed to the reference stations $V_2,\ldots, V_N$ respectively.  The encoded message $u$ is also sent to other stations.  We assume the communication between reference stations are secure, for example QKD system is employed.

\textit{Step 3.} The reference stations $V_1,\ldots,V_N$ agree a time $t_0$ when the PBQC scheme starts.
%They then send their information at time depending on their distance from $P$, in such a way that $P$ is the unique position require shortest time to gather all information.  This places an upper bound to the round-trip time of the process.
At $t=t_0$, $V_1$ sends the encoded qubit to $P$, while $V_i$ sends the classical bit $q_i$ for $i=2,3,\ldots,N$.
%For simplicity, we hereafter consider PBQC starts at $t=0$, and all reference stations located at distance $d$ from $P$, so they send their message at the same time and expect to get the response at $t=2d/c$.

\textit{Step 4.} Upon receiving all information, $P$ adds up all bits to obtain $q$.  The qubit can be decrypted by applying $H^q$, and measured in $Z$ basis to obtain the encoded message.  $P$ broadcasts the results immediately to all reference stations.  We assume all operations of $P$ costs negligible time.

\textit{Step 5.} If $q$'s are random enough, missing any one classical bit would cause half chance of wrong measurement basis.  Reference stations can validate the identity of $P$ by checking if the response is consistent with the encoded message.  By checking the arrival time of the response at different reference stations, location of $P$ is also verified.

%If $q$'s are random enough, then missing any one classical bit would cause half chance of wrong measurement basis.  After receiving the qubit and all information about basis, $P$ conduct measurement immediately and broadcast the result.  Assume all the process of measurement, classical and quantum information manipulation time consume negligible time, the broadcasted result should reach the reference stations no later than a particular time.

%{\it Remark}: Protocol A was claimed to be unconditionally secure and a detailed security proof based on complementarity information tradeoff was given.

\subsection{Protocol B}

The idea of Protocol B \cite{Malaney:2010p5361} is to encode information into maximally entangled states and then encrypted by local transformation.  Information about transformations are sent from different reference stations.  Security of this protocol was supposed to rely on the fact that correct measurement cannot be conducted without decrypting all qubits, as well as the idea that local measurement must disturb an entangled state.  Explicit procedure of Protocol B follows.

\textit{Step 1.} $N$ bits of message is encoded as a $N$ qubit GHZ state
\begin{eqnarray}
|\textrm{GHZ}\rangle &=&  \frac{1}{\sqrt{2}}(|a_1\rangle |a_2\rangle \dots |a_N\rangle \nonumber \\
&&\pm |1\oplus a_1\rangle |1 \oplus a_2\rangle \ldots |1\oplus a_N\rangle)~,
\end{eqnarray}
where $a_1,\ldots,a_N\in \{1,0\}$; $\oplus$ denotes addition with modular 2.  Each reference station picks a qubit from the entangled state.

\textit{Step 2.} Each qubit is encrypted by local transformation $U_i$ and sent to an authorized receiver $P$.  $P$ will store the entangled state in his quantum memory.

\textit{Step 3.} PBQC scheme starts at an agreed time $t=t_0$.  Every reference stations send the classical information about the transformation $U_i$ to $P$ at the same time.

\textit{Step 4.} $P$ immediately decrypts the state after receiving the classical information.  He then conducts a $N$ qubit GHZ state measurement to decode the message, and announces his result at once.

\textit{Step 5.} The measurement result is probably wrong if someone measures the state before getting all transformation information.  Hence the identity of $P$ can be authenticated from the announced result.  Besides, the location of $P$ can be verified by the checking the total time spent in the whole process.

%Each qubit is then encrypted by local transformation $U_i$, and sent to authorized receiver $P$ surrounded by a restricted area that no cheaters can enter.  The PBQC scheme starts when reference stations send classical information about $U$'s to $P$.  Upon the receipt of all $U$'s, $P$ decrypts the states by $\prod_i U^\dag_i$ and conducts multi-partite measurement, the outcome is then broadcasted. The location of $P$ is well chosen that requires the shortest time to collect all messages from reference stations.  Because cheaters outside the restricted area cannot obtain all $U$'s sooner than $P$, cheaters cannot conduct perfect measurement sooner than $P$ even though they have robbed the qubits from $P$.  Reference stations can hence verify the security of communication by checking the response time of the process.

%{\it Remark}: Protocol B was also claimed to be unconditionally secure due to the quantum no-cloning theorem.

\subsection{Dimensionality of PBQC scheme}

In general, the reference stations lie on a one dimensional straight line for $N=2$; a two dimensional plane for $N=3$; and they distribute in the three dimensional space for $N>3$.  We comment that the dimension of position of $P$ that can be verified by PBQC scheme, is \textit{independent} of the spatial dimension of reference station distribution.  For example, even if there are two reference stations and they are collinear with $P$, all three components of $\vec{x}_p=(x_p,y_p,z_p)$ can be authenticated by PBQC scheme.  To illustrate this idea, assume $V_1$ and $V_2$ are lying on the $x$-axis, and $\vec{x}_p$ is some point in between.  The signals are sent at time according to PBQC protocols.  It is easy to see that any position with $y\neq0$ or $z\neq0$ takes a longer time than $P$ to receive both information from $V_1$ and $V_2$.  Thus the \textit{one} dimensional PBQC actually confirms the \textit{three} dimensional position of $P$, instead of the $x$-coordinate only.  Similarly, if $P$ locates at the same plane as three reference stations $V_1,V_2,V_3$, the PBQC scheme also verifies the three dimensional position of $P$.  This argument however requires the position of reference stations are well chosen, i.e. $V_1,V_2,P$ are collinear or $V_1,V_2,V_3,P$ are coplanar.  Four reference stations are necessary if their locations are constrained.

\begin{figure}
\begin{center}
		\includegraphics{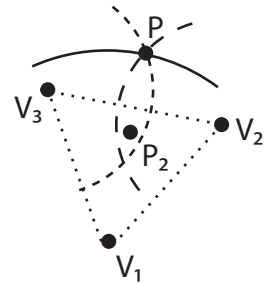}
	\end{center}
	\caption{\label{fig:triangle} At a particular time $t_1$, front of signals sent from $V_1,V_2,V_3$ are represented by solid, dashed, short-dashed lines respectively.  While signals reach $P$ at $t=t_1$, another position $P_2$ inside the triangle of three reference positions (framed by dotted lines) can obtain all information before $t_1$.}
\end{figure}

We also note that PBQC can be performed if and only if there exist a polyhedron formed by positions of some reference stations inside which $P$ locates.  Otherwise for all starting time chosen by the reference stations, there must be places inside the polyhedron such that shorter or equal time is required to receive all information.  The idea is illustrated on Fig.~\ref{fig:triangle}.

\section{Cheating in the $N=2$ case \label{cheat 1D}}

Contrary to claim(s) of unconditional security, we find that both Protocols A and B are, in fact, insecure.
We first demonstrate our cheating strategy for two-reference-stations case (i.e. $N=2$) for both protocols, and generalize it to the more-reference-stations case (i.e., $N>2$) in the next section.  In the current case, we assume $V_1$ and $V_2$ are separated by distance $2d$ and $P$ locates in the middle of two reference stations, so that they lie on a one-dimensional straight line.  PBQC requires $P$ is surrounded by a finite restricted area, such as inside a big military base, with width $2l$ that no cheaters can get into it.  We assume either qubit or classical information are transmitted at the speed of light $c$, and the time for intermediate processing is negligible.  If the PBQC scheme starts at $t=0$, $V_1$ and $V_2$ expect to get the correct response at $t=2d/c$.

\begin{figure}
\begin{center}
\includegraphics{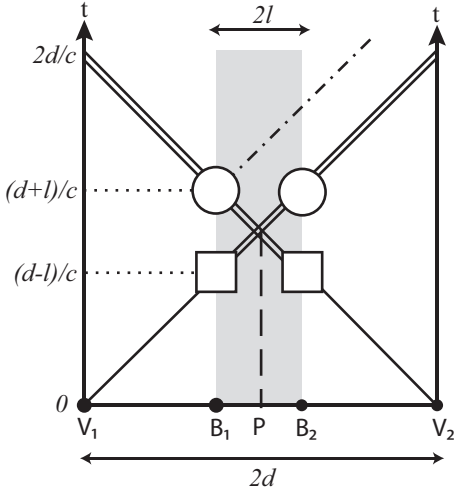}
\caption{Spacetime diagram of the one dimensional scenario.  Solid lines denote spacetime trajectory of information which is possibly quantum or classical, while double lines denote that of classical information only.  In both \textit{Protocols A} and \textit{B}, all measurements ought to be conducted at $t=(d-l)/c$ (shown as squares) in order to give correct response to reference positions on the expected time $t=2d/c$.  Appropriate response is decided after information of another cheater is received at $t=(d+l)/c$ (shown as circles).  If there is no entangled resources shared, $B_1$ has to wait for information from $V_2$ to conduct perfect measurement.  Trajectory of the corresponding response is represented by dot-dashed line, which shows the correct result cannot reach $V_2$ before $t=2d/c$.\label{fig:2_stations}}
\end{center}
\end{figure}

A successful cheating is to produce the correct response not slower than $t=2d/c$ without entering the restricted area.  We find that two cheaters are enough to cheat the protocols in this case.  We assume cheaters $B_1$ and $B_2$ are sitting at $d-l$ and $d+l$ respectively, which are both just outside the restricted area.  The layout of our scenario is shown in Fig.~\ref{fig:2_stations}.

\subsection{Flaw in claimed security proof \label{flaw}}

We remark that Protocol A was once believed to be unconditional secure.
In fact, a detailed claim of proof of security based on \textit{complementary information tradeoff}
was given in \cite{Chandran:2010p5507}.
The intuition behind the claimed proof is that the any measurement on the encrypted qubit
would inevitably disturb the state and hence yield wrong outcome with non-zero probability.

Unfortunately, in the security proof
of Protocol A \cite{Chandran:2010p5507}, it was implicitly assumed that no
prior entanglement is shared by the cheaters. Indeed, a {\it pure} state is assumed for the state consisting of only one cheater and one honest party. See, for example,
in the first sentence of the last paragraph
of p. 8 of the quant-ph version 1 \cite{Chandran:2010p5507}.

We remark that such an assumption is incorrect.
In fact, the cheaters can easily nullify the security proof by using
shared entanglement. We note that with shared entangled resources, quantum teleportation can be conducted by measuring the qubit appropriately \cite{BENNETT:1993p5457}.  The main idea of our cheating scheme in the $N=2$ case is to teleport the encrypted qubit from $B_1$ to $B_2$ for measurement in the correct basis.  Detailed cheating strategy procedure is as follows.

\subsection{Cheating against Protocol A in the $N=2$ case \label{cheat1 1D}}

\textit{Step 1.}Before the cheaters move to the destination, they come together and each pick a particle from a Bell state
\begin{equation}
|\Phi_{00} \rangle \equiv \frac{1}{\sqrt{2}}(|00\rangle + |11\rangle) = \frac{1}{\sqrt{2}}(|++\rangle + |--\rangle) \label{phip}~.
\end{equation}
We assume their quantum memory is perfect that the qubits remain in coherence until measurement.

\textit{Step 2.} When the PBQC scheme starts $t=0$, $V_1$ sends a qubit $H^q |u\rangle$, and $V_2$ sends the bit $q_2=q$ to $P$.  At $t=(d-l)/c$, $B_1$ captures the qubit and $B_2$ obtain the basis information.  To avoid suspicion of $P$, the cheaters can send dummy qubit and basis information to him, and $P$'s response thereafter is interfered or blocked by classical devices.  We hereafter neglect the role of $P$ in our consideration.

\begin{figure}
\begin{center}
		\includegraphics{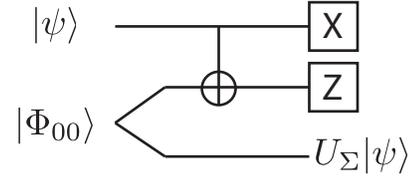}
	\end{center}
	\caption{Circuit for teleporting an unknown qubit $|\psi\rangle$ \cite{ANielsen:2000p5658, BENNETT:1993p5457}.  Measurement is denoted as squares, the measurement basis is represented by the character inside the squares. $U_\Sigma = X^{(1-s_2)/2} Z^{(1-s_1)/2}$ is the byproduct of teleportation depending on random measurement outcome $s_1$ and $s_2$.  \label{fig:teleportation}}
\end{figure}

\textit{Step 3.} $B_1$ immediately perform a Bell measurement on this two qubits in order to teleport the state to $B_2$, the circuit of his measurement is given in Fig.~\ref{fig:teleportation}.  He sends the measurement outcomes of the encrypted qubit, $s_1$, and Bell state qubit, $s_2$, to $B_2$ at once.  We note that measurement outcomes of Pauli operators are $+1$ or $-1$.

\textit{Step 4.}At the same instance $t=(d-l)/c$, the teleported qubit of $B_2$ becomes
\begin{equation}\label{cheat 1D state}
X^{(1-s_2)/2} Z^{(1-s_1)/2} H^q|u\rangle~.
\end{equation}
Consider if $q=0$, $B_2$ has a state
\begin{equation}\label{cheat 1D q0}
X^{(1-s_2)/2} Z^{(1-s_1)/2} |u\rangle = (-1)^{u (1-s_1)/2} |u \oplus (1-s_2)/2\rangle~.
\end{equation}
Since $B_2$ knows the basis is $Z$, and the state in Eq.~(\ref{cheat 1D q0}) is an eigenstate of Pauli $Z$ operator, he can conduct a perfect measurement with outcome $(-1)^{u}s_2$.  Else if $q=0$, $B_2$ has a state
\begin{eqnarray}\label{cheat 1D q1}
&& H Z^{(1-s_2)/2} X^{(1-s_1)/2} |u\rangle \nonumber \\
&=& (-1)^{[u \oplus (1-s_1)/2](1-s_2)/2}H |u \oplus (1-s_1)/2\rangle~.
\end{eqnarray}
Since $B_2$ knows the basis is $X$, and the state in Eq.~(\ref{cheat 1D q1}) is an eigenstate of Pauli $X$ operator, he can conduct a perfect measurement with outcome $(-1)^{u}s_1$.

$B_2$ immediately sends the result to $B_1$.  It is reminded that although the measurement outcome of $B_2$ contains information about outcome of $B_1$ our teleportation scheme does not permit superluminal communication because $B_1$ cannot choose the measurement result deterministically.

\textit{Step 5.} Then at $t=(d+l)/c$, both $B_1$ and $B_2$ knows the result from each other, as well as the correct measurement basis.  They can invert the value of $u$ by multiplying outcome of $B_2$ with second outcome of $B_1$ for $q=0$, or multiplying outcome of $B_2$ with first outcome of $B_1$ for $q=0$.  $B_1$ then sends $u$ to $V_1$ while $B_2$ sends to $V_2$, both reference stations will receive the correct signal at $t=2d/c$.

The whole process consumes the same amount of time to produce the same correct result as there is no cheaters, Protocol A is therefore insecure in 1-D.

{\it Remark}: Let us explain the intuition in Step 3 (the teleportation step). The teleported state received by $B_2$ will be acted upon by one of the four operators, $I, X, Z$ and $ XZ$. Since the original state is an eigenstate of either $X $ and $Z$, we note that the four resulting states are either orthogonal to each other or the same (up to an irrelevant overall phase). Therefore, $B_2$ with the basis information can simply measure the qubit in that basis without disturbing the state at all. Subsequently, after hearing the actual Bell measurement outcome by $B_1$, $B_2$ will be able to tell what the original state is. For this reason, $B_1$ and $B_2$ can cheat successfully with certainty. Therefore, Protocol A is insecure.

{\it Remark}: Let us explain the intuition from another angle.
Since the measurements by $B_1$ and $B_2$ commute with each other, we can also interpret the result from the
viewpoint where $B_2$ performs a measurement before $B_1$ does. In this case, $B_2$, with the basis information will measure a qubit in the correct basis. By the Einstein-Podolsky-Rosen effect, the qubit held by $B_1$ will be projected to either the same state or the opposite state to the qubit sent by $V_1$. So, the task of $B_1$ is to perform a parity check of the states of the two qubits.
While a general parity check is impossible for all bases, we note that in Protocol A, we consider only the two bases $X$ and $Z$.
So, in this case, as the operator $XX$ commutes with $ZZ$, indeed
$B_1$ can perform a parity check by simply doing a Bell measurement.
For this reason, $B_1$ and $B_2$ can cheat successfully with certainty.

%The spacetime diagram of the cheating scheme is shown on Fig.~\ref{fig:2_stations}.  We note that although 2-qubits cluster state is used in our demonstration, teleportation conducted by other maximally entangled state is equally successful.

\subsection{Cheating against Protocol B in the $N=2$ case}

%Consider there are only two reference stations $V_1$ and $V_2$ separated by distance $2d$, and $P$ locates at the middle between them.  $V_1$ and $V_2$ are assumed to be connected by a secure quantum channel, whereby maximally entangled states are shared.  Without loss of generality, the state is assumed to be $|\Phi_{00}\rangle$.

In the current case, 2 bits of information, $ab=\{00,01,10,11\}$, can be encoded into one of the four Bell states $|\Phi_{ab}\rangle$ \cite{Malaney:2010p5361} in Eq.~(\ref{phip}) and
\begin{eqnarray}
|\Phi_{01} \rangle &\equiv& \frac{1}{\sqrt{2}}(|00\rangle - |11\rangle) = \frac{1}{\sqrt{2}}(|+-\rangle + |-+\rangle) \label{phim}\\
|\Phi_{10} \rangle &\equiv& \frac{1}{\sqrt{2}}(|01\rangle + |10\rangle) = \frac{1}{\sqrt{2}}(|++\rangle - |--\rangle)\label{psip}\\
|\Phi_{11} \rangle &\equiv& \frac{1}{\sqrt{2}}(|01\rangle - |10\rangle)= \frac{-1}{\sqrt{2}}(|+-\rangle - |-+\rangle)\label{psim}.
\end{eqnarray}
The qubits are then encrypted by random local transformation $U_1$ and $U_2$ and sent to $P$.  The PBQC scheme starts at $t=0$ when reference stations broadcast $U_i$, and is expected to end at $t=2d/c$ in the honest case.  The idea of the cheating is to first capture and store the qubits until decryption information.  One of the cheater is then teleport the qubit so the other cheater possess two entangled qubit to do Bell measurement.  Step by step procedure is as follows.

\textit{Step 1.} Before the process, the cheaters share a Bell state in Eq.~(\ref{phip}) and store it in good quantum memory.
%We again assume the restricted area has length $2l$ and the cheaters $B_1$ and $B_2$ sit just outside.

\textit{Step 2.} The cheaters break into the quantum channels connecting $P$ with reference stations, they capture the qubits sent by $V_1$ and $V_2$ in a good quantum memory to preserve the coherence until measurement.

\textit{Step 3.} At $t=(d-l)/c$, both cheaters receive the $U_i$ form references, $U^\dag_i$ is applied respectively on the qubits to recover the encoded state $|\Phi_{ab}\rangle$

\textit{Step 4.} $B_2$ teleports the incoming qubit to $B_1$.  We call the encoded state qubit captured by $B_1$($B_2$) as qubit $1$($2$), and the Bell state qubit of $B_1$($B_2$) as qubit $4$($3$).  We analyze the teleportation by stabilizer formalism \cite{Gottesman:1997p4541} as follows.  $B_2$ apply a CNOT gate on his qubits, the state is then stabilized by
\begin{eqnarray}
K_1 = (-1)^a Z_1 Z_2&,&~K_2=(-1)^b X_1 X_2 X_3, \nonumber\\
K_4 = Z_2 Z_3 Z_4&,&~K_3 = X_3 X_4~.
\end{eqnarray}
Qubits $2$ is then measured in $X$ basis and qubit $3$ is measured in $Z$ basis.  The outcomes $s_2$ and $s_3$ are sent to $B_1$ immediately.  New set of stabilizers after the measurement is
\begin{eqnarray}
K_1' = (-1)^a s_3 Z_1 Z_4&,&~K_2' = (-1)^b s_2 X_1 X_4,\nonumber\\
K_3'=s_3 Z_3&,&~K_4'=s_2 X_2~.
\end{eqnarray}
Qubit $2$ and $3$ are obviously no longer entangled as they are measured.  $K_1'$ and $K_2'$ show that qubit $1$ and $4$ are left as a Bell state $|\Phi_{a' b'}\rangle$, where $a'= a+(1-s_3)/2$ and $b' = b+(1-s_2)/2$.  So $B_1$ can measure the state perfectly by Bell measurement, the outcomes $a'$ and $b'$ are sent to $B_2$ immediately.

\textit{Step 5.} At $t=(d+l)/c$, both cheaters obtain information from each others.  $a$ and $b$ are deduced from $a'$, $b'$, $s_2$ and $s_3$, they are sent to and eventually received by reference stations at $t=2d/c$.  Hence correct results are extracted by cheaters using the same time as in honest case, PBQC is hacked.

%The spacetime diagram of the cheating scheme is the same as shown on Fig.~\ref{fig:2_stations}.

\section{Cheating in $N>2$ case \label{cheat 3D}}

%We now consider the cases with $N \geq 3$ reference stations.  In general, the reference stations lie on a two dimensional plane for $N=3$, and they distribute in the three dimensional space in the most general $N>3$ case.
%We comment that the dimension of position of $P$ that can be verified by PBQC scheme, is independent of the spatial dimension of reference station distribution, e.g. even if there are two reference stations and they are lying on a one-dimensional straight line with $P$, all three components of $\vec{x}_p=(x_p,y_p,z_p)$ are authenticated by PBQC scheme.  Using the scenario in the previous section as an illustration.  Assume $V_1,V_2,P$ are lying on the $x$-axis, it is easy to see that any position with $y\neq0$ or $z\neq0$ takes a longer time than $P$ to receive information from both $V_1$ and $V_2$.  Thus the \textit{one} dimensional PBQC actually confirms the \textit{three} dimensional position of $P$, instead of the $x$-coordinate only.  Similarly, if $P$ is coplanar with three reference stations, the PBQC scheme also verifies the three dimensional position of $P$.  This argument however requires the position of reference stations are well chosen, i.e. $V_1,V_2,P$ are collinear or $V_1,V_2,V_3,P$ are coplanar.  Four reference stations are necessary if their locations are constrained.  We also note that PBQC can be performed if and only if there exist a polyhedron formed by positions of some reference stations inside which $P$ locates.

%, which may involve dimension higher than one.

\begin{figure}
\begin{center}
		\includegraphics{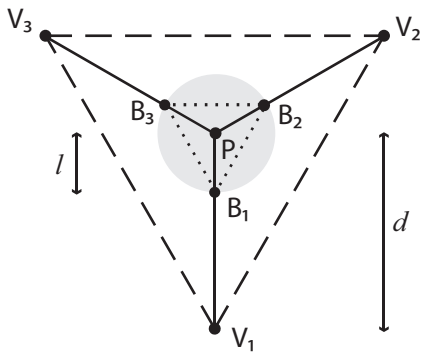}
	\end{center}
	\caption{Positions of reference stations, cheaters, and authorized receiver of our three station scenario are shown as black dots.  The shaded region represents the restricted area surrounding $P$.  Without the cheating, information flows along solid lines; if cheating presents, information flows along solid lines outside the restricted area, and follows dotted lines in the restricted area.  As the path of $V_2 \rightarrow P \rightarrow V_1$ is longer than $V_2 \rightarrow B_2 \rightarrow B_1 \rightarrow V_1$, process of cheating costs shorter time than the honest case.}
	\label{fig:3_stations}
\end{figure}

We first consider the $N=3$ case which the reference stations lie on the same plane, and then discuss how the scheme can be generalized to three dimensional cases with $N>3$.  For simplicity, we assume the three reference stations $V_1, V_2, V_3$ locate at vertex of an equilateral triangle.  The receiver $P$ sits in the center of the triangle, which is distance $d$ from each references, and surrounded by a restricted area with radius $l$.  We find that three cheaters are enough to cheat both protocol perfectly.  We assume cheaters $B_1,B_2,B_3$ locate at distance $l$ from $P$ and $d-l$ from $V_1, V_2, V_3$ respectively.  A layout of their position is shown as Fig.~\ref{fig:3_stations}.

\subsection{Cheating against Protocol A in $N>2$ case}

In this protocol, $V_1$ encrypted the encoded state in $H^q |u\rangle$ and distribute $q_2$ and $q_3$ to $V_2$ and $V_3$.  The physical meaning of $q_2$ and $q_3$ are number of $H$ gate, such that progressively applying $H^{q_2}H^{q_3}$ is equal to $H^q$.  In this case, the cheaters are not going to teleport the qubit as there is only one qubit but two separate piece of information.  Instead they need methods to share the rotation information, so quantum secret sharing scheme is employed \cite{Hillery:1999p5705, Cleve:1999p5862}.  Steps of the cheating scheme is as follows.

\textit{Step 1.} Before the PBQC scheme starts, the cheaters construct a 3-particle GHZ state
\begin{equation}
\frac{1}{\sqrt{2}}(|000\rangle + |111\rangle)~.
\end{equation}
They then travel to the desired position before the PBQC starts at $t=0$.
%\begin{equation}\label{cheat 2D stabilizer}
%K_1=X_1X_2X_3,~K_2=Z_1Z_2,~K_3=Z_1Z_3~,
%\end{equation}
%where the subscripts of Pauli operators are the index of remaining qubits.  We note that the operators in Eq.~\ref{cheat 2D stabilizer} are not the only stabilizers for the state, any product of them, such as $Z_1Z_2=K_2K_3$ is still a stabilizer.
%Each cheater picks a qubit and goes to the assigned position, w.l.o.g. we assume $B_1$ picks the first qubit in the chain.

\textit{Step 2.} At $t=(d-l)/c$, $B_2$ and $B_3$ gets $q_2$ and $q_3$.  If $q_i=0$, cheater $B_i$ measures his qubit in $X$ basis; otherwise he measures in $Y$ basis.  According to the idea of quantum secret sharing, if both $B_2$ and $B_3$ measure in the same basis, the GHZ qubit holding by $B_1$ becomes an eigenstate of Pauli $X$ operator, otherwise it is eigenstate of $Y$ operator.  As an example, let us consider the case $q_2=0$ and $q_3=1$.  It can be shown that the GHZ states before measurement is stabilized by
\begin{equation}
K_1=-Y_1X_2Y_3,~K_2=Z_1Z_2,~K_3=Z_1 Z_3~.
\end{equation}
Entanglement is broken after measurement, and the stabilizers then become single particle operators.  It can be easily obtained that $B_1$'s qubit is then stabilized by $-s_2 s_3 Y$, where $s_2$ and $s_3$ are measurement outcomes of $B_2$ and $B_3$.  Results of other combination of $q$'s are presented in Table~\ref{cheat 2D table}~.

\begin{table}\caption{\label{cheat 2D table} Tables of stabilizers in different cases of $q_i$'s.  $K_1$ is the stabilizer of GHZ state compatible with the measurement basis.  $K_1''$, $K_2''$, and $K_3''$ are stabilizers after the measurement according to the cheating scheme.}
\begin{center}
\begin{tabular}{|c|c|c|c|c|c|}
\hline $q_2$ & $q_3$ & $K_1$ & $K_2''$ & $K_3''$ & $K_1''$ \\
\hline 0 & 0 & $X_1 X_2 X_3$ & $s_2 X_2$ & $s_3 X_3$ & $s_2 s_3 X_1$ \\
\hline 0 & 1 & $-Y_1 X_2 Y_3$ & $s_2 X_2$ & $s_3 Y_3$ & $-s_2 s_3 Y_1$ \\
\hline 1 & 0 & $-Y_1 Y_2 X_3$ & $s_2 Y_2$ & $s_3 X_3$ & $-s_2 s_3 Y_1$ \\
\hline 1 & 1 & $-X_1 Y_2 Y_3$ & $s_2 Y_2$ & $s_3 Y_3$ & $-s_2 s_3 X_1$ \\ \hline
\end{tabular}
\end{center}
\end{table}

\textit{Step 3.} Immediately after the measurement, $B_1$ applies a Hadamard transformation $H$ followed by a $\pi/4$ gate $S$ on the GHZ state qubit, in order to transform the eigenstates of $X$ to that of $Z$, and eigenstates of $Y$ to that of $X$, with the same eigenvalues.  Now it can be observed that if $q_2+q_3$ is even, $B_1$ will have a qubit in $Z$ basis, otherwise he will have a qubit in $X$ basis.  At the the same time $t=(d-l)/c$, $B_1$ also receives the encoded qubit sent from $V_1$, so he has two qubits on hand that are parallel or anti-parallel, i.e. the two qubits are simultaneous eigenstate of either Pauli $X$ or $Z$ operator.  $B_1$ performs Bell measurement and gets one of the four outcomes in Eq.~(\ref{phip}) and (\ref{phim})-(\ref{psim}).

\textit{Step 4.} The cheaters share their measurement outcome and basis information with others.  Since the mutual distance between $B_1$, $B_2$ and $B_3$ is $\sqrt{3}l$, the cheaters can obtain all the information at $t=(d+(\sqrt{3}-1)l)/c$.  From information of $B_2$ and $B_3$, the actual state of the cluster state qubit of $B_1$ is known from Table~\ref{cheat 2D table}.  The outcome of Bell measurement can tell the parity of the two qubits of $B_1$, the state of the encoded qubit is obtained.

\textit{Step 5.} Correct result is then sent by the cheaters and reaches the reference stations at $t=(2d+(\sqrt{3}-2)l)/c$.  When comparing to the case without cheaters that the whole PBQC process is expected to finish at $t=2d/c$, our cheating scheme eventually requires fewer time to produce the correct result.  Cheaters can simply delay for a while before broadcasting their final outcomes, in order to match the time consumption in honest case.  Hence the protocol is cheated.

We note that the time is shortened because information takes $2l/c$ time to travel from $B_1$'s position to $B_2$'s position in honest case, while only $\sqrt{3}l/c$ is needed if there are cheaters.  In general if the geometry is not an equilateral triangle, our cheating scheme may still process faster than the honest case, provided that $P$ is not on the same straight line as any two reference stations.  It is because honest information has to be sent from a vertex to the centre of triangle where $P$ locates, and then rebroadcast to another vertex, while information of cheaters is transmitted along edges of the triangle.

Our cheating scheme can be generalized to cases with $N>3$ reference stations, we need at most $N$ cheaters in each case.  Before the PBQC starts, the cheaters create a $N$ particle GHZ state which is stabilized by
\begin{equation}
K_1=X_1X_2\ldots X_N,~K_i=Z_{i-1} Z_i~,
\end{equation}
for $i=2,3,\ldots,N$; subscripts of Pauli operators are order of remaining qubits.  Each cheater $B_i$ picks a qubit and travels to a position between $P$ and $V_i$.  When $B_2,\ldots,B_N$ receives the basis information, they measure their qubit in $X$ basis if $q_i=0$, or $Y$ basis if $q_i=1$, and broadcast the results.  If even number of $q$'s are equal to 1, the qubit of $B_1$ is in $X$ basis, otherwise it changes to eigenstate of $Y$ basis.  In the former case, $Y$ measurement must be performed in pairs.  Consider $q_i=1$ at position $m$ and $n$, we must be able to construct a stabilizer $K_1' = K_1 K_{m+1} K_{m+2} \ldots K_{n}= -X_1\ldots Y_m \dots Y_n \dots$ which is compatible to the measurements, such that the qubit of $B_1$ remains at $X$ basis after the measurement.  Otherwise there is one single $Y$ measurement at position $r$, the compatible stabilizer becomes $K_1'=K_1 K_2\ldots K_{r}=-Y_1\dots Y_r \dots$, and the qubit of $B_1$ has changed to eigenstate of $Y$ basis.

Identical to the $N=3$ case, $B_1$ applies an $SH$ gate onto his cluster state qubit, he then obtains an eigenstate of $X$ operator if $q$ is odd or an eigenstate of $Z$ operator if $q$ is even.  He then measures the cluster state qubit and encoded qubit sent from $V_1$ by Bell measurement and broadcasts the measurement outcome.  In the present case of $N>3$, cheaters do not receive all information at the same time, but it is easy to check that even the slowest piece of information should arrive as late as in the honest case.  Information provided by $B_2,\ldots,B_N$ determines the actual state of GHZ state qubit of $B_1$, and the measurement of $B_1$ reveals the parity between his two qubits.  Hence the value of encoded qubit is obtained and cheaters sent the results to reference stations.  The whole process takes fewer time than or the same as the honest case.

\subsection{Cheating against Protocol B in $N>2$ case}

In this protocol, 3 bits of information is encoded as one of the eight tripartite GHZ-type states \cite{Malaney:2010p5361} characterizing by parameters $b_1, b_2, b_3$
\begin{equation}
|\Phi_{b_1 b_2 b_3}\rangle = \frac{1}{\sqrt{2}}(|a_1\rangle |a_2\rangle |a_3\rangle \pm |1\oplus a_1\rangle |1 \oplus a_2\rangle |1\oplus a_3\rangle )~,
\end{equation}
where $a_1, a_2, a_3, b_1, b_2, b_3 \in \{0,1\}$; $(-1)^{b_1}=\pm1$ is the phase between two superposition states; $b_2=a_1\oplus a_2$ and $b_3=a_1\oplus a_3$.  The quibits are then distributed to reference stations, we denote the qubit held by $V_i$ as qubit $i$.
%It is noted that the encoding require manipulations of at least two of the three qubits, it ought to be conducted before distribution, in differ from the case with two reference stations only.
Qubit $i$ is encrypted by arbitrary local transformation $U_i$, and sent to $P$ subsequently.  The PBQC scheme starts at $t=0$ when reference stations send information of $U_i$ to $P$, correct response should return at $t=2d/c$.  We find that three cheaters are enough to cheat perfectly in this case.  The idea is the same as in $N=2$ case, which is to teleport all qubits to one cheater, so he can conduct a $N$ particle GHZ-type state measurement.  The cheating strategy is as follows.

\textit{Step 1.} Before they have travelled to desire positions, $B_2$($B_1$) picks qubit $4$ ($5$), and $B_3$($B_1$) picks qubit $6$($7$), which qubits $4$ and $5$, $6$ and $7$ are Bell states in Eq.~(\ref{phip}).

\textit{Step 2.} The cheaters break into the quantum channel connecting $V_i$ and $P$ and capture the encrypted qubits.  We call the qubit captured by $B_1$ as qubit $i$.  At $t=(d-l)/c$, cheaters receive information of $U_i$ and corresponding decryption procedure is made to obtain the original encoded state.

\textit{Step 3.} Immediately after the decryption, $B_2$ and $B_3$ conducts Bell measurement for teleportation.  Afterwards, the state is stabilized by
\begin{eqnarray}
K_1' &=& (-1)^{b_1} s_2 s_3 X_1 X_5 X_7, \nonumber \\
K_2' &=& (-1)^{b_2} s_4 Z_1 Z_5, ~K_3'= (-1)^{b_3}s_6 Z_1 Z_7, \nonumber \\
K_4' &=& s_2 X_2,~K_5'=s_4 X_4 \nonumber \\
K_6' &=& s_3 X_3,~K_7'=s_6 X_6~. \label{cheat2 2D}
\end{eqnarray}
It is easy to verify from Eq.~(\ref{cheat2 2D}) that qubits $1$, $5$, $7$ becomes a GHZ-type state $|\Phi_{b_1' a_2' a_3'}\rangle$, where $b_1' = b_1 \oplus (1-s_2 s_3)/2$, $b_2' = b_2  \oplus (1-s_4)/2$, $b_3'= b_3  \oplus (1-s_5)/2$.  So $B_1$ can conduct GHZ state measurement to reveal the residual state exactly.  Result is sent to other cheaters.

\textit{Step 4.} In our equilateral triangle case, information exchange among cheaters is finished at $t=(d+(\sqrt{3}-1)l)/c$.  The encoded message $b_1$, $b_2$, $b_3$ can easily be inferred from the measuring outcomes.  Correct results are sent to reference stations, and the whole process can be finished as early as $t=(2d+(\sqrt{3}-2)l)/c$, which is even shorter than the honest case.  If the three stations are not forming an equilateral triangle, the time required by cheating is longer.  But the time consumption is in general fewer than $2d/c$ for any 3-station scenario, PBQC is hence cheated.

\section{Principle of the cheating schemes \label{cheating principle}}

\subsection{Protocol A}

We have verified that our cheating strategy works not only for qubits encoded in BB84 states (eigenstates of $X$ and $Z$ operators), it also works if eigenstates of both Pauli $X$, $Y$, and $Z$ can be chosen for encoding.  In the one-dimensional case, the same cheating scheme can be applied as described in Section \ref{cheat1 1D}. What is the intuition here? An intuition is that the teleported state will be transformed by one of the four operators $I, X, Z,$ and $XZ$. Now, if the input state is an eigenstate of either $X$, or $Y$ or $Z$ operator, then the output state will be either the same or opposite to the input state (up to an irrelevant overall phase).
Another intuition is that a Bell measurement by $B_1$ allows him to check the parity of operators, $XX$, $YY$ and $ZZ$ simultaneously
as the three operators commute with each other.

  In the case of more than 3 reference stations, the original quantum secret sharing idea does not work, because we need switching between the three basis instead of two in Protocol A.  Cheating can be achieved by cluster state quantum computation (CSQC) \cite{Raussendorf:2003p2988}.  Instead of a $N$-particle GHZ state, the cheaters shares a $4N-3$ particles chain cluster state.  $B_1$ picks a qubit on the end of the chain, while other cheaters pick 4 consecutive qubits from the chain.  As stated in \cite{Raussendorf:2003p2988}, each cheater can conduct general rotation by measuring three qubits in appropriate direction, while the last qubit is measured in $X$ basis to teleport the state to next cheater.  Finally, the cluster state qubit of $B_1$ lies in the same basis as the incoming qubit, and he can conduct Bell measurement as before.  It is noted that all cluster state qubit measurements can be conducted at the same time, sequence of the measurement among cheaters is unimportant.  It is because all measurement are local operations and obviously independent to each other.  We also note that the quantum secret sharing and quantum teleportation mentioned before are special cases of CSQC.  In fact CSQC is a more general concept, so we will analyze our cheating scheme under this formalism.

We find that two characteristics of eigenstates in $X$, $Y$, and $Z$ basis leave the possibility for our cheating.  Firstly, the conversion between them ($H$ is the transformation between $X$ states and $Z$ states; $S$ is the transformation between $X$ states and $Y$ states) are in the Clifford group.  Recall in the $N=2$ case, we have pulled the $H$ gate from Eq.~(\ref{cheat 1D state}) to the front in Eq.~(\ref{cheat 1D q1}), and let the Pauli operators applying on the $Z$ states before the gate.  Since the basis of Pauli states are not changed by Pauli operators, the action of transformation gate is not altered and hence the teleported state is in the same basis as the original qubit.  In $N>2$ cases, we refer to the general rotation operation of CSQC \cite{Raussendorf:2003p2988},
\begin{equation}\label{cheat general state}
|\psi_{\textrm{out}}\rangle = \prod_{i=2}^N U_{\Sigma_i} U_i |\psi_{\textrm{in}}\rangle~,
\end{equation}
where $|\psi_{out}\rangle$ is the cluster state qubit held by $B_1$; $|\psi_{\textrm{in}}\rangle$ can be treated as $|0\rangle$ in our case; $U_i$ is rotation induced by cluster state measurement, in our case it is performed by $B_i$ to conduct rotation hinted by the message of $V_i$; $U_{\Sigma_i}=X^i Z^i$ is the by product of random measurement outcome of the $i$-th qubit.  It is transparent that if $U_i$ are all operators in the Clifford group, Eq.~(\ref{cheat general state}) becomes
\begin{equation}\label{cheat Clifford state}
|\psi_{\textrm{out}}\rangle = U U_{\Sigma} |0\rangle = e^{i \phi} U |0~\textrm{or}~1\rangle~,
\end{equation}
where $U$ is the product of all $U_i$, which is the complete basis information separated beforehand in this protocol; $U_{Sigma}$ is a product of Pauli operators, its form depends $U_{\Sigma_i}$ as well as commutating relation between $U_{\Sigma_i}$ and $U_i$; $e^{i \phi}=\{\pm1, \pm i \}$ is the phase generated by $U_{\Sigma} |0\rangle$, and the state $|0\rangle$ can only flip to $|1\rangle$ or remain unchanged upon $U_{\Sigma}$.

The cluster state qubit of $B_1$ is hence parallel or anti-parallel to the encoded qubit, $B_1$ can obtain information of the unknown qubit by checking parity of his qubits on hand, if such parity checking measurement exists.  For eigenstates of Pauli operators, parity of two qubits in the same basis can be checked by Bell measurement, which is the second key point to our cheating scheme.  We illustrate the idea using the BB84 states and leave interested readers to verify the $Y$ states.  If both qubits of $B_1$ are in $Z$ basis, only $|\Phi_{00}\rangle$ and $|\Phi_{01}\rangle$ contains even parity states and odd parity states appear in $|\Phi_{10}\rangle$ and $|\Phi_{11}\rangle$ only; while if they are in $X$ basis, only $|\Phi_{00}\rangle$ and $|\Phi_{10}\rangle$ contains even parity states and odd parity states appear in $|\Phi_{01}\rangle$ and $|\Phi_{11}\rangle$ only.  It can be seen that even and parity states do not appear in the same Bell states, hence cheaters can infer the parity of qubits of $B_1$ by the Bell measurement result.  Furthermore, cheaters know the exact form of cluster state qubit of $B_1$, the qubit is then revealed by the parity.

\subsection{Protocol B}

This protocol is once believed to be secure.  The argument is based on the quantum no-cloning theorem \cite{Malaney:2010p5361}.  But it is not necessary to clone the state perfectly in order to conduct a perfect measurement distant apart.  The problem of Protocol B is that the message is encoded in GHZ states, which each code is related to each other by single bit flip and overall phase flip only.  Since the random byproduct of our teleportation scheme are single particle $X$ and $Z$ operators, an encoded state must be mapped to another code state after teleportation.  As a result, a standard decoding procedure can read out the teleported state perfectly.

\section{Modified PBQC scheme \label{proposal}}

In Section \ref{cheating principle}, we have discussed Protocol A is insecure because the qubit is encoded as eigenstates of $X$ or $Z$, such that the basis of state do not change upon teleportation or cluster state manipulation.  And we are able to reveal the parity of two particles if they are in $X$ or $Z$ basis.
Notice, however, that if one modifies a protocol to allow more general bases other than the $X$, $Y$ and $Z$ bases,
then our cheating strategy does not generally work, in the sense that $B_1$ may not be able to find an appropriate basis to measure the teleported qubits perfectly. This is because the byproduct of random measurement outcomes of teleportation may map the encoded state into a state that is no longer a code.  In fact, our cheating scheme fails if references encoded the 2-bit message as the states $\{|00\rangle, (|01\rangle \pm |10\rangle)/\sqrt{2}, |11\rangle\}$ in the one dimensional case.  It is easy to check that the by products of teleportation do not necessarily map a code state to a code state.  But if the cheaters are allowed to possess general entangled resources, is this protocol still secure?  We do not have the answer right now, and leave it as an open question as our next goal of endeavour.

A natural modification to Protocol A hopefully resistant to the cheating is to encode the message as $\pm1$ eigenstates of $\hat{n}(\theta,\phi)\cdot \vec{\sigma}$
\begin{eqnarray}\label{general psi}
|\psi\rangle &=& \cos \frac{\theta}{2} |0\rangle + \sin \frac{\theta}{2} e^{i \phi} |1\rangle~, \\
|\bar{\psi}\rangle &=&\sin \frac{\theta}{2} |0\rangle - \cos \frac{\theta}{2} e^{i \phi} |1\rangle~,
\end{eqnarray}
where $0 \leq \theta \leq \pi$ and $0 \leq \phi \leq 2 \pi$, $\vec{\sigma}=X\hat{x}+Y\hat{y}+Z\hat{z}$ is the Pauli vector.
Such a modified protocol was also proposed in \cite{Kent:2010p5759}.

In the one dimensional case, reference stations $V_1$ and $V_2$ are assumed connecting with a quantum channel, through which QKD can be conducted.  Although only finite bits can be communicated through QKD, random $\theta$ and $\phi$ can be generated by various methods.  One example is to make use of the universality of quantum computation, that arbitrary qubit can be constructed by sequences of Clifford operators plus a $\pi/8$ gate \cite{ANielsen:2000p5658, Malaney:2010p5361}.  A random sequence of `0' and `1' is generated by $V_1$ and sent to $V_2$.  Each `0' represents an operation of Hadamard gate $H$, while a `1' represents an operation of $\pi/8$ gate $T$, such that the encrypted state is given by $HTHTT |u\rangle$, where $u\in \{0,1\}$ is the bit to be encoded, for sequence `01011' \cite{Malaney:2010p5361}.  Similiarly if there are $N$ reference stations, each of them are connected to $V_1$ with QKD channels for communication of arbitrary rotation $U_i$.  The encrypted qubit sent from $V_1$ to $P$ is $U_2 \ldots U_N |u\rangle$, while reference stations $V_2 \ldots V_N$ send information of rotation $U_i$.

It is not difficult to check that our modified protocol is immune to our original cheating strategy
demonstrated in Section \ref{cheat 1D}
 and \ref{cheat 3D}.  In the one dimensional case, $B_1$ captures the state $|\psi\rangle$ in Eq.~(\ref{general psi}) sent from $V_1$ at $t=(d-l)/c$, and teleported to $B_2$ immediately.  Although $B_2$ knows the basis from $V_2$, it can be shown that the teleported state can be neither parallel nor anti-parallel with the original one, i.e. for $s_1$ and $s_2$ not equal to one, the matrix element
\begin{equation}
\langle \psi | X^{(1-s_2)/2} Z^{(1-s_1)/2} |\psi \rangle \neq 0~\textrm{or}~1~.
\end{equation}
Therefore $B_2$ cannot find a basis to measure the qubit always perfectly without knowing measurement outcomes from $B_1$.  Message of $B_1$ arrives $B_2$ as soon as $t=(d+l)/c$.  Even if $B_2$ measures the qubit immediately, correct feedback will reach $V_1$ no earlier than $t=2(d+l)/c$ which costs more time than expected.  Security of PBQC is hence enforced.

In the case of more reference stations, rotation $U_i$'s do not belong to the Clifford group, it precludes the encrypted state to transform from Eq.~(\ref{cheat general state}) to Eq.~(\ref{cheat Clifford state}).  As in the one dimensional case, $B_1$ cannot perform a perfect measurement until the random outcomes of cluster state measurement are known.  But the location where all information can reach each other in the shortest time locates inside the restricted area of $P$, in other words the cheaters need time than the honest case to get the correct result.  Security of PBQC is hence enforced.

In practice, neither the quantum operations, quantum channel, nor measurements are noiseless, incorrect response can be given even in the honest case.  The total error rate of practical PBQC system ought to be bounded below the successful cheating rate, i.e. probability of producing correct feedback by cheaters on time.  Otherwise failure of cheating may be regarded as error caused by noise, PBQC scheme hence becomes insecure.

We now discuss the successful cheating rate of our protocol under various cheating schemes in the one dimensional case.  First of all, we consider $B_1$ simply measures the qubit and announces the result.  It is equal to the average value of $|\langle 0 | \psi\rangle|^2$ for any $\theta$ and $\phi$.  The successful rate is obviously $50\%$, as it is no different from a random guess.  Next, we consider $B_1$ measures the qubit but announces until obtaining basis information from $B_2$.  The successful rate is $75\%$.  It is more than a random guess, because for $\theta<\pi/2$, $B_1$'s measurement outcome is more probably correct, while it is more probably wrong if $\theta>\pi/2$.  $B_1$ can announce the inverse of his measurement outcome for $\theta>\pi/2$ case, successful rate is then increased.  Finally we consider our teleportation cheating scheme.  $B_2$ measures the teleported state in Fig.~\ref{fig:teleportation} by basis states $\{|\psi\rangle,|\bar{\psi}\rangle\}$.  Consider if the result is $|\psi\rangle$.  After knowing $s_2$ and $s_3$, the cheaters announce the more probably correct result, i.e.
\begin{equation}
|v\rangle = \{ |\psi\rangle, |\bar{\psi}\rangle | \max(|\langle \psi| X^{(1-s_2)/2} Z^{(1-s_1)/2}|v\rangle|^2) \}~.
\end{equation}
The average successful rate is
\begin{eqnarray}
&&\frac{1}{4} \int \Big[ 1+\max(|\langle \psi| X|\psi\rangle|^2,|\langle \psi| X|\bar{\psi}\rangle|^2) \nonumber \\
&&+\max(|\langle \psi| Z|\psi\rangle|^2,|\langle \psi| Z|\bar{\psi}\rangle|^2) \nonumber \\
&&+\max(|\langle \psi| XZ|\psi\rangle|^2,|\langle \psi| XZ|\bar{\psi}\rangle|^2) \Big] d\Omega~,
\end{eqnarray}
which is about $85\%$.  We have checked numerically that $85\%$ is the highest successful rate can be achieved for any measurement basis is used by $B_2$.  In the case of more reference stations, cluster state quantum computation requires more measurements.  So there are more random byproducts and the successful rate is anticipated to be lower than one dimensional case.

It is noted that Chandran \textit{et al.} suggests $B_1$ can entangle the encoded qubit with his quantum memory, and then sends one to $B_2$.  Since the quantum systems are entangled, any measurement of $B_1$ will leave $B_2$ a mixed state.  $B_1$ has to announce a response before knowing measurement result of $B_2$, otherwise the operation time must exceed that allowed by PBQC scheme.  But measurement outcome of $B_2$ is probabilistic, if $B_2$ makes a response according to this, there is probability that the response received by $V_1$ and $V_2$ are inconsistent.  This kind of inconsistency reveals there must be cheaters lying between, because noisy operation in honest case must not produce inconsistent results.  So either $B_1$ or $B_2$ must not do any measurement, the cheaters are not benefited by the quantum memory.

\section{Security of modified protocol \label{security}}

So far we have demonstrated how our modified PBQC protocol remains secure against the teleportation-based cheating scheme.  It is curious to know whether the protocol is secure if other kinds of entangled qubits are shared among cheaters, and strategy other than teleportation is employed.

Recalling in the one dimensional case, we first teleport the unknown encoded state from $B_1$ to $B_2$ for measurement; while in $N>2$ case we use secret sharing ideas to send a share of the basis
information to $B_1$ through cluster state quantum computation, parity of the cluster state qubit and encoded qubit is then checked.  Although seemingly different cheating strategy is taken in one dimensional and $N>2$ cases, they are actually the same in principle.  It is because the measurements of cheaters are local and thus independent on each others.  Time order of the measurement is unimportant, all cheaters conduct their operation immediately after receiving information reference stations.  It is easy to see that measurement of cluster state by $B_2$ in the one dimensional case actually uses secret sharing ideas to send a share of the basis information to $B_1$, while the entanglement operation and $X$ basis measurement of $B_1$ is equivalent to some parity checking procedure.

We first consider the one dimensional case.  Suppose a general 2-qubit entangled state is shared among cheaters, which would become any set of states containing basis information after measurements of $B_2$.  If the encoded qubit is in $Z$ basis, we assume w.l.o.g. that $B_2$ makes an measurement to `send' states $|\tilde{0}\rangle$ and $|\tilde{1}\rangle$ to $B_1$.  In general, the states $B_1$ obtained from $B_2$ are
\begin{eqnarray}
|\tilde{\uparrow}\rangle &=& g(\theta,\phi) |\tilde{0}\rangle + h(\theta,\phi)|\tilde{1}\rangle \nonumber \\
|\tilde{\downarrow}\rangle &=& h^\ast(\theta,\phi) |\tilde{0}\rangle -g^\ast(\theta,\phi)|\tilde{1}\rangle~,
\end{eqnarray}
where $g,h$ are functions of basis information of $B_2$.  Here we have assumed $|\tilde{\uparrow}\rangle$ and $|\tilde{\downarrow}\rangle$ are orthogonal, successful rate of cheating decreases in non-orthogonal case.  It is noted that the probability for $|\tilde{\uparrow}\rangle$ or $|\tilde{\downarrow}\rangle$ to appear equals to $0.5$, otherwise causality is violated.  After the measurement, $B_2$ transmits classical information of basis of encoded qubit, basis of his measurement on the entangled state, and the measurement outcome to $B_1$.

In the $Z$-basis case, in order to distinguish $|0\rangle$ and $|1\rangle$ after obtaining information from $B_2$, $B_1$ conducts a von Neumann measurement with basis
\begin{eqnarray}
|M_1\rangle &=& \alpha |0\rangle |\tilde{0}\rangle + \beta |1\rangle |\tilde{1}\rangle,~|M_2\rangle = \beta^\ast |0\rangle |\tilde{0}\rangle - \alpha^\ast |1\rangle |\tilde{1}\rangle \nonumber \\
|M_3\rangle &=& \gamma |0\rangle |\tilde{1}\rangle + \delta |1\rangle |\tilde{0}\rangle,~|M_4\rangle = \delta^\ast |0\rangle |\tilde{1}\rangle - \gamma^\ast |1\rangle |\tilde{0}\rangle~, \nonumber \\ \label{B_1 measurement}
\end{eqnarray}
where the coefficients $\alpha,\beta,\gamma,\delta$ characterize the measurement.  The states are set as above such that every $|\tilde{0}\rangle$ and $|\tilde{1}\rangle$ associate with either $|0\rangle$ or $|1\rangle$.  Otherwise if $|M_i\rangle$ contains terms like $(|0\rangle + |1\rangle)|\tilde{1}\rangle$, $B_1$ cannot reveal the identity of encoded qubit after communicated with $B_2$.

Since $B_1$ knows nothing about the basis, he always conducts the same measurement in Eq.~(\ref{B_1 measurement}).  For general $\theta$ and $\phi$, $B_1$ gets one of the four states
\begin{equation}\label{B_1 states}
\{|\psi\rangle |\tilde{\uparrow}\rangle, |\psi\rangle |\tilde{\downarrow}\rangle, |\bar{\psi}\rangle |\tilde{\uparrow}\rangle, |\bar{\psi}\rangle |\tilde{\downarrow}\rangle\}~.
\end{equation}
An important observation is that the cheaters are able to distinguish the encoded qubit, only if every measurement state $|M_i\rangle$ contains components of $|\tilde{\uparrow}\rangle$ and $|\tilde{\downarrow}\rangle$ associated with either $|\psi\rangle$ or $|\bar{\psi}\rangle$ only.  This statement can be reformulated to say that each state in Eq.~(\ref{B_1 states}) is a superposition of at most two states in Eq.~(\ref{B_1 measurement}).

We expand $|\psi\rangle |\tilde{\uparrow}\rangle$ as
\begin{eqnarray}
|\psi\rangle |\tilde{\uparrow}\rangle &=& \cos\frac{\theta}{2} g(\theta,\phi) |0\rangle |\tilde{0}\rangle + \cos\frac{\theta}{2} h(\theta,\phi) |0\rangle |\tilde{1}\rangle \\
&&+ \sin\frac{\theta}{2} e^{i \phi} g(\theta,\phi) |1\rangle |\tilde{0}\rangle + \sin\frac{\theta}{2} e^{i \phi} h(\theta,\phi) |1\rangle |\tilde{1}\rangle~. \nonumber
\end{eqnarray}
W.l.o.g. we assume it is superposition of $|M_1\rangle$ and $|M_3\rangle$, imposing the relations
\begin{equation}
\cot\frac{\theta}{2} e^{-i \phi} \frac{g}{h} = \frac{\alpha}{\beta},~\cot\frac{\theta}{2} e^{-i \phi} \frac{h}{g} = \frac{\gamma}{\delta}~.
\end{equation}
Similarly we expand $|\bar{\psi}\rangle |\tilde{\downarrow}\rangle$
\begin{eqnarray}\label{up up}
|\bar{\psi}\rangle |\tilde{\downarrow}\rangle &=& \sin\frac{\theta}{2} h^\ast(\theta,\phi) |0\rangle |\tilde{0}\rangle - \sin\frac{\theta}{2} g^\ast(\theta,\phi) |0\rangle |\tilde{1}\rangle \\
&&- \cos\frac{\theta}{2} e^{i \phi} h^\ast(\theta,\phi) |1\rangle |\tilde{0}\rangle + \cos\frac{\theta}{2} e^{i \phi} g^\ast(\theta,\phi) |1\rangle |\tilde{1}\rangle~. \nonumber
\end{eqnarray}
It has to be superposition of either $|M_1\rangle$ and $|M_3\rangle$ or $|M_2\rangle$ and $|M_4\rangle$, otherwise unphysical result $\langle \psi \tilde{\uparrow} | \bar{\psi} \tilde{\downarrow}\rangle \neq 0$ is yielded.

We first consider $ |\bar{\psi} \tilde{\downarrow}\rangle$ is a superposition of $|M_2\rangle$ and $|M_4\rangle$.  The following relations has to be satisfied
\begin{equation}\label{down down}
\cot\theta e^{-i \phi} \frac{g}{h} = -\frac{\alpha}{\beta},~\cot\theta e^{-i \phi} \frac{h}{g} = -\frac{\gamma}{\delta}~.
\end{equation}
Together with Eq.~(\ref{up up}), $\alpha, \delta, g(\theta, \phi)$ ought to be zero.  This implies $B_2$ always sends $|\tilde{0}\rangle$ and $|\tilde{1}\rangle$ to $B_1$, and the basis of $B_1$ measurement is four untangled states, i.e. two single qubit measurement.  It can be readily seen that both $|\psi\rangle |\tilde{0}\rangle$ and $|\bar{\psi}\rangle |\tilde{0}\rangle$ contain component of $|M_2\rangle$, hence the cheaters cannot distinguish $|\psi\rangle$ and $|\bar{\psi}\rangle$ after communication.

Now we consider $ |\bar{\psi} \tilde{\downarrow}\rangle$ is a superposition of $|M_1\rangle$ and $|M_3\rangle$, imposing the relations
\begin{equation}\label{down down 2}
\tan\frac{\theta}{2} e^{-i \phi} \frac{h^\ast}{g^\ast} = \frac{\alpha}{\beta},~\tan\frac{\theta}{2} e^{-i \phi} \frac{g^\ast}{h^\ast} = \frac{\gamma}{\delta}~.
\end{equation}
Together with Eq.~(\ref{up up}), we have
\begin{equation}
\frac{|g|^2}{|h|^2}=\tan^2\frac{\theta}{2},~\frac{|h|^2}{|g|^2}=\tan^2\frac{\theta}{2}~,
\end{equation}
which can only be satisfied for $\theta=\pi/2$ but not for general $\theta$.  We therefore conclude that our protocol is unbreakable no matter what 2-qubit state is shared among cheaters, and what strategy the cheaters employ.  We would like to comment on the case of $\theta=\pi/2$ case, it means that the basis for encoding are perpendicular in the Bloch's sphere.  There are only three mutually perpendicular directions in the Bloch's sphere, which can be regarded as the $X$, $Y$ and $Z$ directions.  It explains why our cheating works for and only for states encoded in eigenstates of Pauli $X$, $Y$, $Z$ operators.

In the case with $N>2$ reference stations, we can prove by contradiction that our scheme is secure for any $N$-qubit states is shared by the cheaters and any strategy they take.  Let the maximum mutual distance between reference stations is $2d$, we call the two maximally separated reference stations $V_1$ and $V_2$.  Assume $P$ lies in the middle such that the minimum time required for PBQC scheme is $t=2d/c$.  Suppose the restricted area of $P$ is so large that the $N$ cheaters have to sit very close to each reference stations.  The minimum time required for information exchange among cheaters is $t=2d/c$.  If there exists a strategy to cheat successfully for any $\theta,\phi$, and any $U_i$ distributed to $V_i$.  Consider the case that $U_3,\ldots,U_N$ are identity operators, in other words the basis information is contained in $U_2$ only.  After measurement of $B_3,\ldots,B_N$, the situation reduces to the one dimensional case, which cannot be cheated as proved above.  Therefore perfectly successful strategy for $N>2$ reference stations case does not exist.

In general, the cheaters can share more complicated quantum resources than entangled qubits.  If two qubits belonging to an entangled network, such as a 2D cluster state, are possessed by each cheaters, it has to be treated as an entangled four-level system which is not covered in our proof.  A PBQC protocol is unconditional secure if and only if cheating cannot succeed with certainty no matter what entangled resources is shared among cheaters, and what strategy they take.  We provide in Appendix \ref{security 3 level} the proof that our protocol is still secure if entangled three-level system is shared among cheaters.  We believe our protocol is still secure in general, because $B_2$'s measurement on high-level entangled system should produce more random outcomes, which is not beneficial to cheating.  But we cannot prove it analytically right now and leave it as an open question.

\section{Summary \label{conclusion}}

We have shown how entangled resources can help cheating the two proposed PBQC protocols.  The idea is to use teleportation, quantum secret sharing or CSQC to share part of quantum information among cheaters, whereby measurement can be conducted before all information is known.  Subsequent exchange of classical information amends the result to compensate the random measurement outcomes of entangled resources.

Our cheating scheme is successful because random byproducts of teleportation, quantum secret sharing or CSQC, are Pauli operators.  They do not map any code state out of the code space.  The loophole can be fixed by using non-Clifford states to encode the message.  Based on this idea, we propose a modified version of Protocol A which the code space spans the Bloch's sphere.  Our protocol is proved to be secure if each cheater share entangled qubits or qutrit (such as particle composites with effective spin 1).  The highest average successful rate of cheating our protocol is $75\%$ if no entangled resource is used, and $85\%$ if our cheating scheme is employed.

After the completion and circulation of a preliminary version of the
current paper, we have learnt of a preprint by Buhrman \textit{et al.} \cite{Buhrman:2010p6512}, which claims that,
for a rather general class of protocols, if cheaters possess unlimited amount of entanglement, then all PBQC protocols are insecure.
If their argument is correct, then it will be interesting to know whether PBQC protocols remains secure
if the cheaters only share finite amount of entanglement, which is more realistic, and what the minimum amount of entanglement is required to cheat a PBQC scheme perfectly.  Another intriguing question is whether PBQC is still feasible if some secret is shared by authorized receiver and reference stations, instead of communicating all information through public channels.  Very recently, Kent \cite{Kent:2010p6513} has proposed a PBQC scheme that is claimed to be secure, provided that $P$ and reference stations agree with a sequence of bits that cannot be obtained by cheaters.  Besides, Malaney has also considered a protocol that entangled pairs are beforehand shared among $P$ and reference stations \cite{Malaney:2010p5555}.

Finally, we have assumed in the above consideration that all operations are performed extremely fast comparing to the traveling time of signals.  To verify the possibility of PBQC in practice, we consider the distance between reference stations is at the order of $100$km, and the size of restricted area of $P$ is order of km.  A round trip of signals takes around $100~\mu$s, and the presence of cheating will give a deviation for about $1\%$ of time, which is $\mu$s scale.  Consider recent experiments on optical quantum computation are operating on nanosecond scales \cite{Prevedel:2007p6161}, the assumption of fast operation is still valid.

We thank Marcos Curty for bringing the subject of position-based quantum cryptography to our attention. His enlightening discussions started our investigation. We also thank numerous colleagues including Eric Chitambar,
Harry Buhrman, Serge Fehr, Andrian Kent, Bill Munro, Renato Renner and Barry Sanders for useful conversations. We thank funding agencies and programs including NSERC, CIFAR, QuantumWorks, and Canada Research Chair program for financial support.

%However teleportation is just one example of quantum phenomena making use of entanglement, it is curious to know the if the protocol is unconditionally secure, that is no matter what kind of entangled resource is shared among cheaters, and what kind of strategy is performed by cheaters, correct response cannot be produced faster than or as fast as the honest case.

%Unfortunately, we do not have adequate analytical tool to prove the unconditional security.

\appendix
\section{Security of shared 3-level system\label{security 3 level}}

Here we outline the proof of security of our protocol if an entangled three-level system is shared among two cheaters in one dimensional case.  This guarantee the security in cases involving $N>2$ reference stations, of which the one dimensional case can always be regarded as a special case, as claimed in the Section \ref{security}.

First of all, we investigate the properties of the entangled $n$-level system if the cheating is successful with certainty.  We again assume $B_2$ can make a measurement such that $B_1$ will receive one of $n$ orthogonal states with equal probability.  The choice of orthogonal states incorporate information about the basis of encoded qubit.  If the encoded qubit is an eigenstate of $Z$, we let $B_1$ will receive a states belonging to $\{|\phi_1\rangle,\ldots,|\phi_n\rangle \}$.  Define a vector $(|\vec{\phi}\rangle)\equiv (|\phi_1\rangle \dots |\phi_n\rangle)^{T}$, such that if the encoded qubit is an eigenstate of $\hat{n}(\theta,\phi)\cdot\vec{\sigma}$, $B_1$ will receive an element of the vector $\hat{T}(\theta,\phi)(|\vec{\phi}\rangle)$, where $\hat{T}(\theta,\phi)$ is an $n\times n$ unitary matrix freely chosen by $B_2$.

Let the basis of $B_1$'s measurement operator is $\{|M_1\rangle,\ldots,|M_n\rangle \}$.  In order to distinguish the identity of encoded qubit after exchanging information with $B_2$, each state $|M_i\rangle$ should not contain components of both $|\phi_j\rangle|0\rangle$ and $|\phi_j\rangle|1\rangle$ for any $j$.  Define a selection matrix $S^{(i)}$ which is an diagonal matrix with eigenvalues $1$ and $0$ only, such that $S^{(i)}_jj=1$ if $|M_i\rangle$ contains component of $|\phi_j\rangle|0\rangle$; while $S^{(i)}_jj=0$ if $|M_i\rangle$ contains component of $|\phi_j\rangle|1\rangle$.  The states $|M_i\rangle $ can be written as
\begin{equation}\label{M Z}
|M_i\rangle = \sum_{j,j'} \left(\alpha_{ij'} S^{(i)}_{j'j} |\phi_j\rangle |0\rangle + \alpha_{ij'} (I-S^{(i)}_{j'j}) |\phi_j\rangle |1\rangle\right)~,
\end{equation}
where $|\alpha_{i1}|^2+\ldots+|\alpha_{in}|^2=1$; $I$ is the $n\times n$ identity matrix.  Since $B_1$ knows nothing about the basis, his measurement is always the same.  Similar to the argument above, if $B_1$ is able to distinguish $|\psi\rangle$ and $|\bar{\psi}\rangle$ after information exchange, $|M_i\rangle$ has to be
\begin{eqnarray}\label{M n}
|M_i\rangle &=& \sum_{k,k',j} \Big(\beta_{ik} \tilde{S}^{(i)}_{kk'} \hat{T}_{k'j} |\phi_j\rangle |\psi\rangle \nonumber \\
&&+ \beta_{ik} (I-\tilde{S}^{(i)}_{kk'}) \hat{T}_{k'j}|\phi_j\rangle |\bar{\psi}\rangle\Big)~,
\end{eqnarray}
where $|\beta_{i1}|^2+\ldots+|\beta_{in}|^2=1$; $\tilde{S}^{(i)}$ are selection matrices.  Comparing Eq.s~(\ref{M Z}) and (\ref{M n}), we get
\begin{eqnarray}
\alpha_{ij} [ \cos\frac{\theta}{2} S^{(i)}_{jj} + \sin\frac{\theta}{2} e^{i \phi} (I-S^{(i)}_{jj}) ] &= &\beta_{ik} \tilde{S}^{(i)}_{kk} \hat{T}_{kj} \\
\alpha_{ij} [ \sin\frac{\theta}{2} S^{(i)}_{jj} - \cos\frac{\theta}{2} e^{i \phi} (I-S^{(i)}_{jj}) ] &=& \beta_{ik} (I-\tilde{S}^{(i)}_{kk}) \hat{T}_{kj}~. \nonumber \\
\end{eqnarray}
Summing the above two relations, and taking the scalar product of themselves, we have
\begin{eqnarray}
&&\sum_{j} \alpha_{ij} [ (1+\sin\theta) S^{(i)}_{jj} + (1-\sin\theta) (I-S^{(i)}_{jj})] \alpha^\ast_{ij} \nonumber \\
&=& \sum_k \beta_{ik} \beta^\ast_{ik} = 1~,
\end{eqnarray}
where we have made use of the identities $S^2=S$, $(I-S)^2=(I-S)$, $S(I-S)=0$.  For $\sin\theta \neq 0$, above equation  together with normalization of $|M_i\rangle$ imposes
\begin{equation}\label{cheat even}
\sum_{j} \alpha_{ij} S^{(i)}_{jj}\alpha^\ast_{ij} = \sum_{j} \alpha_{ij}(I-S^{(i)}_{jj}) \alpha^\ast_{ij} =\frac{1}{2}~.
\end{equation}
This is an important relation to restrict the kind of measurement conducted by $B_1$.  It is to say that if cheating is successful, the absolute square sum of coefficients associated with $|0\rangle$ of any state $|M_i\rangle$ has to be $\frac{1}{2}$.

Here we show that a complete set of states satisfying Eq.~(\ref{cheat even}) does not exist in the entangled three-level system case.  Assume $|M_1\rangle$ consists of $\{|0\rangle|\phi_1\rangle, |0\rangle|\phi_2\rangle, |1\rangle|\phi_3\rangle \}$ with non-zero contribution.  Other $|M\rangle$ cannot be formed by combination with two $|0\rangle$ and one $|1\rangle$, otherwise the inner product with $|M_1\rangle$ is non-zero, unless coefficient of one $|0\rangle$ is  zero.  We ignore this case for a while and assume other $|M\rangle$'s has to be formed by one $|0\rangle$ and two $|1\rangle$.  Consider $|M_2\rangle$ share two common components as $|M_1\rangle$, for example it contains $\{|0\rangle|\phi_1\rangle, |1\rangle|\phi_2\rangle, |1\rangle|\phi_3\rangle \}$.  By completeness relation of the measurement states, there must be a state $|M_3\rangle$ contains $|1\rangle|\phi_1\rangle$.  But there is no state containing $|1\rangle|\phi_1\rangle$ can both orthogonal to $|M_1\rangle$, $|M_2\rangle$ and satisfy Eq.~(\ref{cheat even}).  So $|M_2\rangle$ must contain $\{|1\rangle|\phi_1\rangle, |1\rangle|\phi_2\rangle, |0\rangle|\phi_3\rangle \}$.  Since there are totally six orthogonal states, at least three of them must share the same set of components.

Consider $|M_1\rangle,|M_3\rangle,|M_5\rangle$ contain the terms $\{|0\rangle|\phi_1\rangle, |0\rangle|\phi_2\rangle, |1\rangle|\phi_3\rangle \}$.  As they have to satisfy Eq.~(\ref{cheat even}), they can be expressed as
\begin{equation}
|M_i\rangle = \frac{1}{\sqrt{2}}\left(\cos\theta_i |0\rangle|\phi_1\rangle + \sin\theta_i e^{i\mu_i}|0\rangle|\phi_2\rangle+ e^{i \nu_i}|1\rangle|\phi_3\rangle \right)~.
\end{equation}
Since the three states are orthogonal, we require
\begin{equation}
\cos\theta_i \cos\theta_j +\sin\theta_i \sin\theta_j e^{i (\mu_i - \mu_j)} = - e^{i(\nu_i - \nu_j)}~,
\end{equation}
for $i\neq j$.  The term on the right hand side has norm $1$, which imposes constraints on left hand side such that
\begin{equation}
|M_3\rangle = \frac{1}{\sqrt{2}}\left(\cos\theta_1 |0\rangle|\phi_1\rangle + \sin\theta_1 e^{i\mu_1}|0\rangle|\phi_2\rangle- e^{i \nu_1}|1\rangle|\phi_3\rangle \right)~.
\end{equation}
But one cannot find $(\theta_5, \mu_5, \nu_5)$ that $|M_5\rangle$ is orthogonal to $|M_1\rangle$ and $|M_3\rangle$.  So there does not exist three states sharing the same components of states.  We now return to the case that some $|M_i\rangle$ contains only two $|0\rangle$ and one $|1\rangle$ with coefficient of one $|0\rangle$ is zero.  As the coefficient is zero, it is no different to treat the component as $|1\rangle$, it will then fall into paradigm of our proof.

We have assumed in the above argument that at least one state is a superposition of three components, we now consider all states contain only two components.  We find the only possible choice of states is
\begin{eqnarray}\label{3 states 1}
|M_{1,2}\rangle&=&\frac{1}{\sqrt{2}} \left( |0\rangle |\phi_1\rangle \pm e^{i \mu_1}|1\rangle |\phi_2\rangle \right) \\
&=&\frac{1}{\sqrt{2}} |\psi\rangle \left(  \cos\frac{\theta}{2}|\phi_1\rangle \pm \sin\frac{\theta}{2}e^{i (\mu_1-\phi)} |\phi_2\rangle \right) \nonumber \\
&&+\frac{1}{\sqrt{2}} |\bar{\psi}\rangle \left( \sin\frac{\theta}{2}|\phi_1\rangle \mp \cos\frac{\theta}{2}e^{i (\mu_1-\phi)} |\phi_2\rangle \right) \nonumber \\
|M_{3,4}\rangle&=&\frac{1}{\sqrt{2}} \left( |0\rangle |\phi_2\rangle \pm e^{i \mu_2}|1\rangle |\phi_3\rangle \right) \\
&=&\frac{1}{\sqrt{2}} |\psi\rangle \left(  \cos\frac{\theta}{2}|\phi_2\rangle \pm \sin\frac{\theta}{2}e^{i (\mu_2-\phi)} |\phi_3\rangle \right) \nonumber \\
&&+\frac{1}{\sqrt{2}} |\bar{\psi}\rangle \left( \sin\frac{\theta}{2}|\phi_2\rangle \mp \cos\frac{\theta}{2}e^{i (\mu_2-\phi)} |\phi_3\rangle \right) \nonumber \\ \label{3 states 3}
|M_{5,6}\rangle&=&\frac{1}{\sqrt{2}} \left( |0\rangle |\phi_3\rangle \pm e^{i \mu_3}|1\rangle |\phi_1\rangle \right) \\
&=&\frac{1}{\sqrt{2}} |\psi\rangle \left(  \cos\frac{\theta}{2}|\phi_3\rangle \pm \sin\frac{\theta}{2}e^{i (\mu_3-\phi)} |\phi_1\rangle \right) \nonumber \\
&&+\frac{1}{\sqrt{2}} |\bar{\psi}\rangle \left( \sin\frac{\theta}{2}|\phi_3\rangle \mp \cos\frac{\theta}{2}e^{i (\mu_3-\phi)} |\phi_1\rangle \right) \nonumber ~,
\end{eqnarray}
or some cyclic permutation of $|\phi\rangle$.  On the other hand, because we have proved $|M\rangle$ cannot contains three components, each $|M_i\rangle$ should be written as,
\begin{eqnarray}\label{3 states 4}
|M_{1,2}\rangle&=&\frac{1}{\sqrt{2}} \left( |\psi\rangle (\hat{T}_{1i}|\phi_i\rangle) \pm e^{i \nu_1}|\bar{\psi}\rangle (\hat{T}_{2i}|\phi_i\rangle) \right) \\
|M_{3,4}\rangle&=&\frac{1}{\sqrt{2}} \left( |\psi\rangle (\hat{T}_{2i}|\phi_i\rangle) \pm e^{i \nu_1}|\bar{\psi}\rangle (\hat{T}_{3i}|\phi_i\rangle) \right) \\ \label{3 states 6}
|M_{5,6}\rangle&=&\frac{1}{\sqrt{2}} \left( |\psi\rangle (\hat{T}_{3i}|\phi_i\rangle) \pm e^{i \nu_1}|\bar{\psi}\rangle (\hat{T}_{1i}|\phi_i\rangle) \right)~.
\end{eqnarray}
When comparing the state associated with $|\psi\rangle$ in $|M_1\rangle$ and $|M_3\rangle$, it is clear that the expressions in Eq.~(\ref{3 states 1})-(\ref{3 states 3}) and Eq.~(\ref{3 states 4})-(\ref{3 states 6}) cannot be equivalent.  Therefore, we now can claim that $B_1$ cannot find a measurement such that the state $|\psi\rangle$ and $|\bar{\psi}\rangle$ after exchanging information with $B_2$, no matter what set of 3-level states is transmitted by $B_2$ through the entangled resources, which proves security of our protocol in this case.

\bibliographystyle{phaip}
\pagestyle{plain}
\bibliography{PBC}

\end{document}